\documentclass[letterpaper]{article}
\usepackage{aaai}
\usepackage{times}
\usepackage{helvet}
\usepackage{courier}
\usepackage{multirow}
\usepackage{array}
\usepackage{adjustbox}
\usepackage{subfigure}
\usepackage{color}
\usepackage{threeparttable}

\title{Is the Sample Good Enough? Comparing Data from Twitter's Streaming API with Twitter's Firehose}
\author{Fred Morstatter\\
  Arizona State University\\
  699 S. Mill Ave.\\
  Tempe, AZ, 85281\\
\And
J{\"u}rgen Pfeffer\\
  Carnegie Mellon University\\
  5000 Forbes Ave.\\
  Pittsburgh, PA, 15213\\
\And
Huan Liu\\
  Arizona State University\\
  699 S. Mill Ave.\\
  Tempe, AZ, 85281\\
\And
Kathleen M. Carley\\
  Carnegie Mellon University\\
  5000 Forbes Ave.\\
  Pittsburgh, PA, 15213\\
}

\begin{document}
\maketitle

\begin{abstract}
\begin{quote}
Twitter is a social media giant famous for the exchange of short, 140-character messages called ``tweets''. In the scientific community, the microblogging site is known for openness in sharing its data. It provides a glance into its millions of users and billions of tweets through a ``Streaming API'' which provides a sample of all tweets matching some parameters preset by the API user. The API service has been used by many researchers, companies, and governmental institutions that want to extract knowledge in accordance with a diverse array of questions pertaining to social media. The essential drawback of the Twitter API is the lack of documentation concerning what and how much data users get. This leads researchers to question whether the sampled data is a valid representation of the overall activity on Twitter. In this work we embark on answering this question by comparing data collected using Twitter's sampled API service with data collected using the full, albeit costly, Firehose stream that includes every single published tweet. We compare both datasets using common statistical metrics as well as metrics that allow us to compare topics, networks, and locations of tweets. The results of our work will help researchers and practitioners understand the implications of using the Streaming API.
\end{quote}
\end{abstract}

\begin{figure*}[t]
     \begin{center}
        \subfigure[Firehose]{%
           \label{fig:tcfire}
           \includegraphics[width=0.5\textwidth]{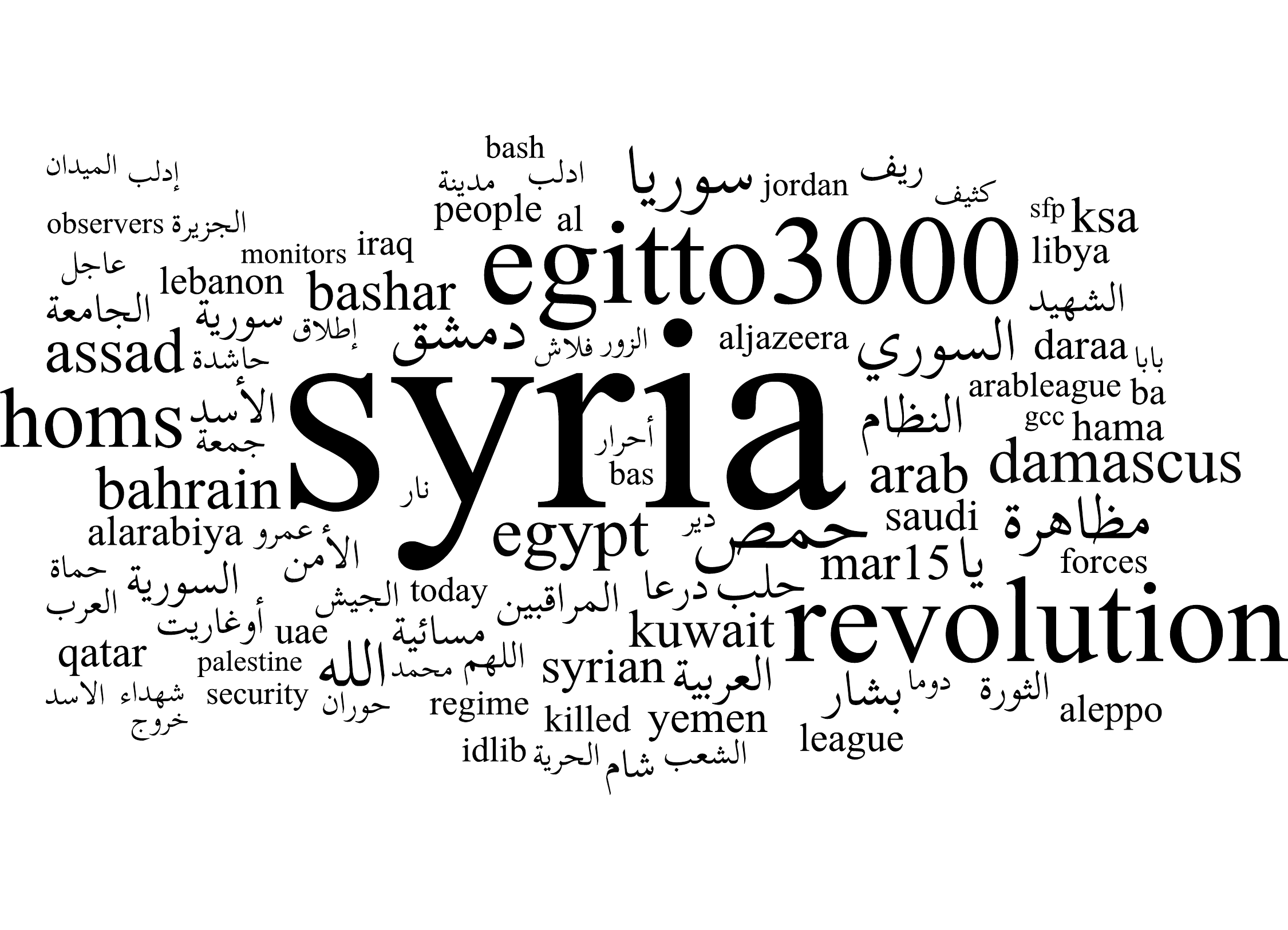}
        }~ %
        \subfigure[Streaming API]{
            \label{fig:tcstream}
            \includegraphics[width=0.5\textwidth]{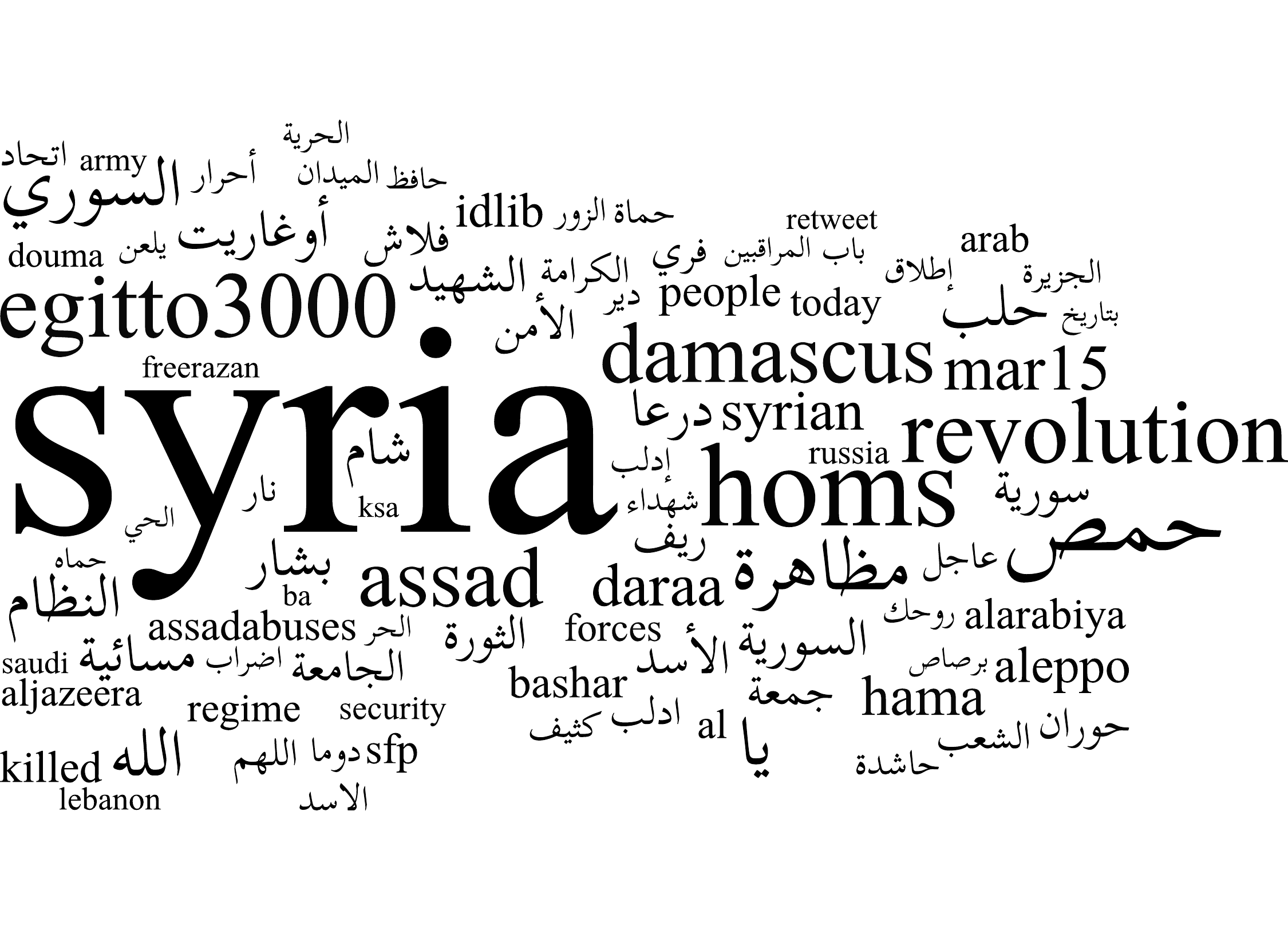}
        }%
    \end{center}
    \caption{Tag cloud of top terms from each dataset.}
   \label{fig:tagclouds}
\end{figure*}
\section{Introduction}
Twitter is a microblogging site where users exchange short, 140-character messages called ``tweets''. Ranking as the 10th most popular site in the world by the Alexa rank in January of 2013\footnote{http://www.alexa.com/topsites}, the site boasts 500 million registered users publishing 400 million tweets per day. Twitter's platform for rapid communication is said to be a vital communication platform in recent events including Hurricane Sandy\footnote{http://www.nytimes.com/interactive/2012/10/28/nyregion/hurricane-sandy.html}, the Arab Spring of 2011~\cite{Camp2011}, and several political campaigns~\cite{tuma10,gay11}. As a result, Twitter's data has been coveted by both computer and social scientists to better understand human behavior and dynamics.

Social media data is often difficult to obtain, with most social media sites restricting access to their data. Twitter's policies lie opposite to this. The ``Twitter Streaming API''\footnote{https://dev.twitter.com/docs/streaming-apis} is a capability provided by Twitter that allows anyone to retrieve at most a 1\% sample of all the data by providing some parameters. According to the documentation, the sample will return at most 1\% of all the tweets produced on Twitter at a given time. Once the number of tweets matching the given parameters eclipses 1\% of all the tweets on Twitter, Twitter will begin to sample the data returned to the user. The methods that Twitter employs to sample this data is currently unknown. The Streaming API takes three parameters: keywords (words, phrases, or hashtags), geographical boundary boxes, and user ID. 

One way to overcome the 1\% limitation is to use the Twitter Firehose---a feed provided by Twitter that allows access to 100\% of all public tweets. A very substantial drawback of the Firehose data is the restrictive cost. Another drawback is the sheer amount of resources required to retain the Firehose data (servers, network availability, and disk space). Consequently, researchers as well as decision makers in companies and government institutions are forced to decide between two versions of the API: the freely-available but limited Streaming, and the very expensive but comprehensive Firehose version. To the best of our knowledge, no research has been done to assist those researchers and decision makers by answering the following: How does the use of the Streaming API affect common measures and metrics performed on the data? In this article we answer this question from different perspectives. 

We begin the analysis by employing classic statistical measures commonly used to compare two sets of data. Based on unique characteristics of tweets, we design and conduct additional comparative analysis. By extracting topics using a frequently used algorithm, we compare how topics differ between the two datasets. As tweets are linked data, we perform network measures of the two datasets. Because tweets can be geo-tagged, we compare the geographical distribution of geolocated tweets to better understand how sampling affects aggregated geographic information.

\section{Related Work}
Twitter's Streaming API has been used throughout the domain of social media and network analysis to generate understanding of how users behave on these platforms. It has been used to collect data for topic modeling~\cite{Hong-Davi10,Pozd-11}, network analysis~\cite{Sofe12}, and statistical analysis of content~\cite{Math10}, among others. Researchers' reliance upon this data source is significant, and these examples only provide a cursory glance at the tip of the iceberg. Due to the widespread use of Twitter's Streaming API in various scientific fields, it is important that we understand how using a sub-sample of the data generated affects these results.

From a statistical point of view, the ``law of large numbers'' (mean of a sample converges to the mean of the entire \emph{population})
and the Glivenko-Cantelli theorem (the unknown distribution $X$ of an attribute in a population can be approximated with the observed distribution $x$) guarantee satisfactory results from sampled data when the randomly selected sub-sample is big enough. From network algorithmic~\cite{Wasserman1994} perspective the question is more complicated. Previous efforts have delved into the topic of network sampling and how working with a restricted set of data can affect common network measures. The problem was studied earlier in~\cite{gran76}, where the author proposes an algorithm to sample networks in a way that allows one to estimate basic network properties. More recently,~\cite{Cost03} and~\cite{Borg06} have studied the affect of data error on common network centrality measures by randomly deleting and adding nodes and edges. The authors discover that centrality measures are usually most resilient on dense networks. In~\cite{Koss06}, the authors study global properties of simulated random graphs to better understand data error in social networks.~\cite{leskovec2006sampling} proposes a strategy for sampling large graphs to preserve network measures. 

In this work we compare the datasets by analyzing facets commonly used in the literature. We start by comparing the top hashtags found in the tweets, a feature of the text commonly used for analysis. In~\cite{tsur2012s}, the authors try to predict the magnitude of the number of tweets mentioning a particular hashtag. Using a regression model trained with features extracted from the text, the authors find that the content of the idea behind the tag is vital to the count of the tweets employing it. Tweeting a hashtag automatically adds a tweet to a page showing tweets published by other tweeters containing that hashtag. In~\cite{Yang-etal12}, the authors find that this communal property of hashtags along with the meaning of the tag itself drive the adoption of hashtags on Twitter.~\cite{de2010does} studies the propagation patterns of URLs on sampled Twitter data.

Topic analysis can also be used to better understand the content of tweets.~\cite{kire09} drills the problem down to disaster-related tweets, discovering two main types of topics: informational and emotional. Finally,~\cite{Yin11,Hong12,Pozd-11} all study the problem of identifying topics in geographical Twitter datasets, proposing models to extract topics relevant to different geographical areas in the data.~\cite{Jose12} studies how the topics users discuss drive their geolocation.

Geolocation has become a prominent area in the study of social media data. In~\cite{Waka11} the authors try to classify towns based upon the content of the geotagged tweets that originate from within the town.~\cite{Delo09} studies Twitter's use as a sensor for disaster information by studying the geographical properties of users tweets. The authors discover that Twitter's information is accurate in the later stages of a crisis for information dissemination and retrieval.

\section{The Data}
\begin{figure}[t]
\includegraphics[width=0.48\textwidth]{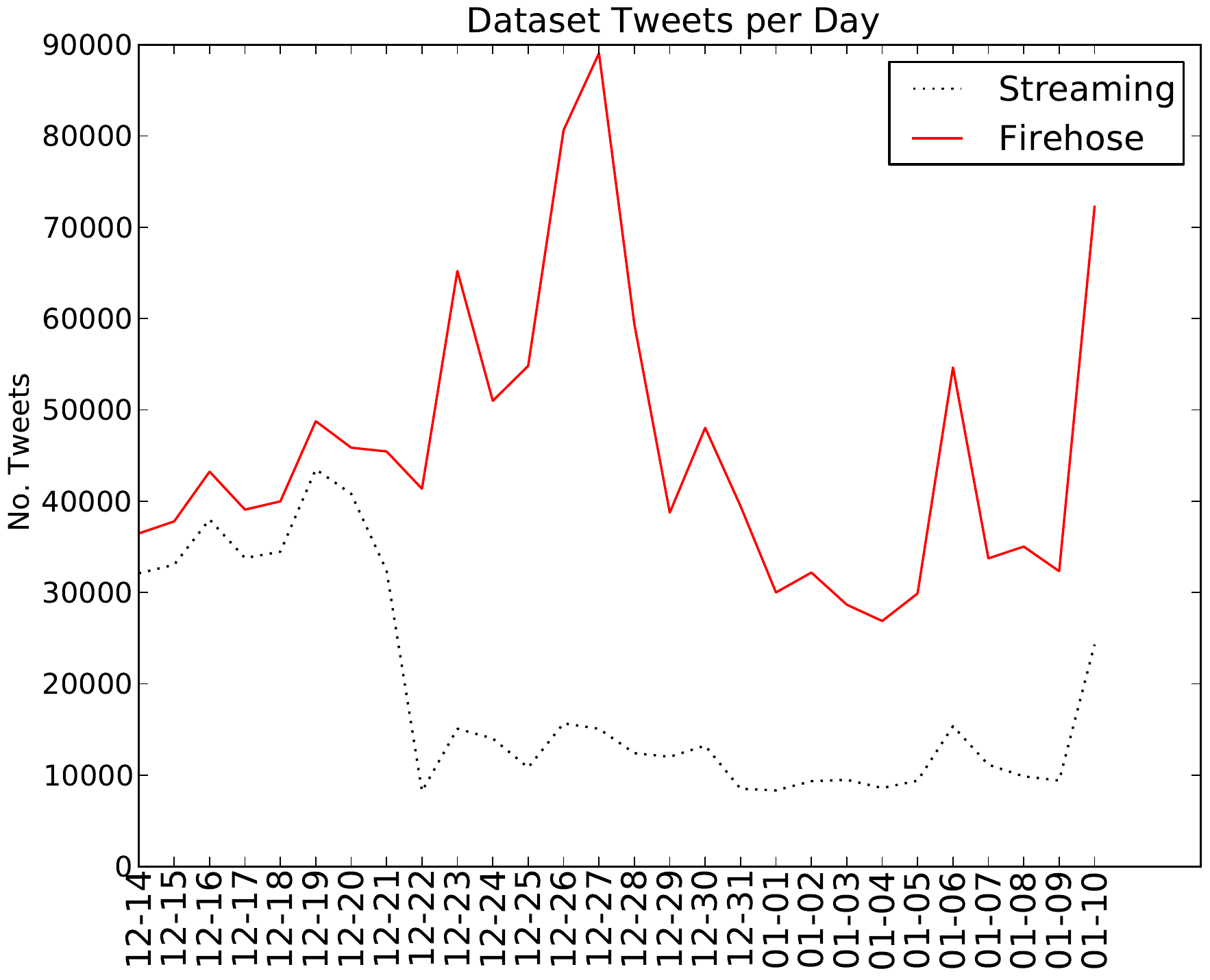}
\caption{Raw tweet counts for each day from both the Streaming API and the Firehose.}
\label{fig:counts}
\end{figure}

From December 14th, 2011 - January 10th, 2012 we collected tweets from the Twitter Firehose matching any of the keywords, geographical bounding boxes, and users in Table~\ref{tab:params}. During the same time period, we collected tweets from the Streaming API using TweetTracker~\cite{Kumar2011} with exactly the same parameters. During the time we collected 528,592 tweets from the Streaming API and 1,280,344 tweets from the Firehose. The raw counts of tweets we received each day from both sources are shown in Figure~\ref{fig:counts}. One of the more interesting results in this dataset is that as the data in the Firehose spikes, the Streaming API coverage is reduced. One possible explanation for this phenomenon could be that due to the Western holidays observed at this time, activity on Twitter may have reduced causing the 1\% threshold to go down.

\begin{figure}[t]
  \centering
  \includegraphics[width=0.3\textwidth]{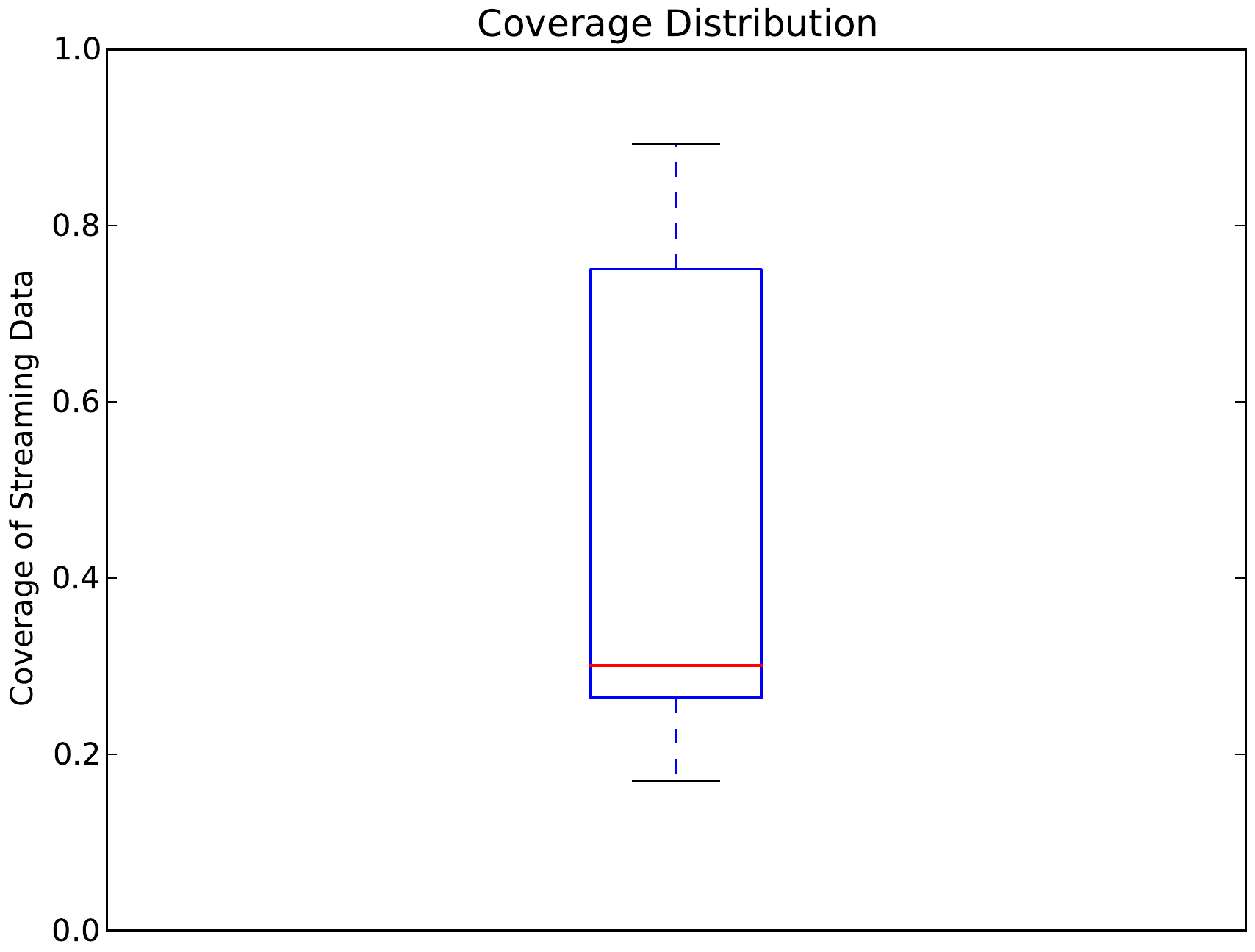}
  \caption{Distribution of coverage for the Streaming data by day. Whiskers indicate extreme values.}
  \label{fig:covbox}
\end{figure}

One of the key questions we ask in this work is how the amount of coverage affects measures commonly performed on Twitter data. Here we define coverage as the ratio of data from the Streaming API to data from the Firehose. To better understand the coverage of the Streaming API for each day, we construct a box-and-whisker plot to visualize the distribution of daily coverage, shown in Figure~\ref{fig:covbox}. In this period of time the Streaming API receives, on average, 43.5\% of the data available on the Firehose on any given day. While this is much better than just 1\% of the tweets promised by the Streaming API, we have no reference point for the data in the tweets we received.

The most striking observation is the range of coverage rates (see Figure~\ref{fig:covbox}). Increase of \emph{absolute} importance (more global awareness) or \emph{relative} importance (the overall number of tweets decreases) result in lower coverage as well as fewer tweets. To give the reader a sense for the top words in both datasets, we include tag clouds for the top words in the Streaming API and the Firehose, shown in Figure~\ref{fig:tagclouds}.

\begin{table}[t]
  \caption{Parameters used to collect data from Syria. Coordinates below the boundary box indicate the Southwest and Northeast corner, respectively.}
  \begin{threeparttable}
  \small \begin{tabular}{| p{2.6cm} | c | c |}
  \hline
  Keywords & Geoboxes & Users \\
  \hline
  \#syria, \#assad, \#aleppovolcano, \#alawite, \#homs, \#hama, \#tartous, \#idlib, \#damascus, \#daraa, \#aleppo, \#\includegraphics[scale=0.2]{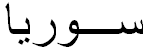}\tnote{*}, \#houla 
  &
  \raisebox{-\totalheight}{
    \includegraphics[scale=0.2]{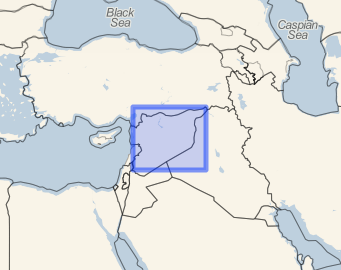}
  }
  &
  @SyrianRevo \\
  & (32.8, 35.9), (37.3, 42.3) & \\
  \hline
  \hline
  \end{tabular}
  \begin{tablenotes}
    \item [*] Arabic word for ``Syria''
  \end{tablenotes}
  \label{tab:params}
  \end{threeparttable}
\end{table}

\section{Statistical Measures}
We investigate the statistical properties of the two datasets with the intent of understanding how well the characteristics of the sampled data match those of the Firehose. We begin first by comparing the top hashtags in the tweets for different levels of coverage using a rank correlation statistic. We continue to extract topics from the text, matching topical content and comparing topical distribution to better understand how sampling affects the results of this common process performed on Twitter data. In both cases we compare our streaming data to random datasets obtained by sampling the data obtained through the Firehose. 

\subsection{Top Hashtag Analysis}
Hashtags are an important communication device on Twitter. Users employ them to annotate the content they produce, allowing for other users to find their tweets and to facilitate interaction on the platform. Also, adding a hashtag to a tweet is equivalent to joining a community of users discussing the same topic~\cite{Yang-etal12}. In addition, hashtags are also used by Twitter to calculate the trending topics of the day, which encourages the user to post in these communities. 

Recently, hashtags have become an important part of Twitter analysis~\cite{efron2010hashtag,tsur2012s,Recu12}. For both the purpose of community formation and trend analysis it is important that our Streaming dataset convey the same importance for hashtags as the Firehose data. Here we compare the top hashtags in the two datasets using Kendall's $\tau$ rank correlation coefficient~\cite{agre10}.
\subsubsection{Kendall's $\tau$ of Top Hashtags}
\label{sec:kthash}
Kendall's $\tau$ is a statistic which measures the correlation of two ordered lists by analyzing the number of concordant pairs between them. Consider two hashtags, \#A and \#B. If both lists rank \#A higher than \#B, then this is considered a concordant pair, otherwise it is counted as a discordant pair. Ties are handled using the $\tau_\beta$ statistic as follows:
\begin{equation}
\tau_\beta = \frac{|P_C| - |P_D|}{\sqrt{(|P_C| + |P_D| + |T_F|)(|P_C| + |P_D| + |T_S|)}}
\end{equation}
where $P_C$ is the set of concordant pairs, $P_D$ is the set of discordant pairs, $T_F$ is the set of ties in the Firehose data, but not in the Streaming data, $T_S$ is the number of ties found in the Streaming data, but not in the Firehose, and $n$ is the number of pairs in total. The $\tau_\beta$ value ranges from -1,  perfect negative correlation, to 1, perfect positive correlation.

\begin{figure}[t]
  \includegraphics[width=0.5\textwidth]{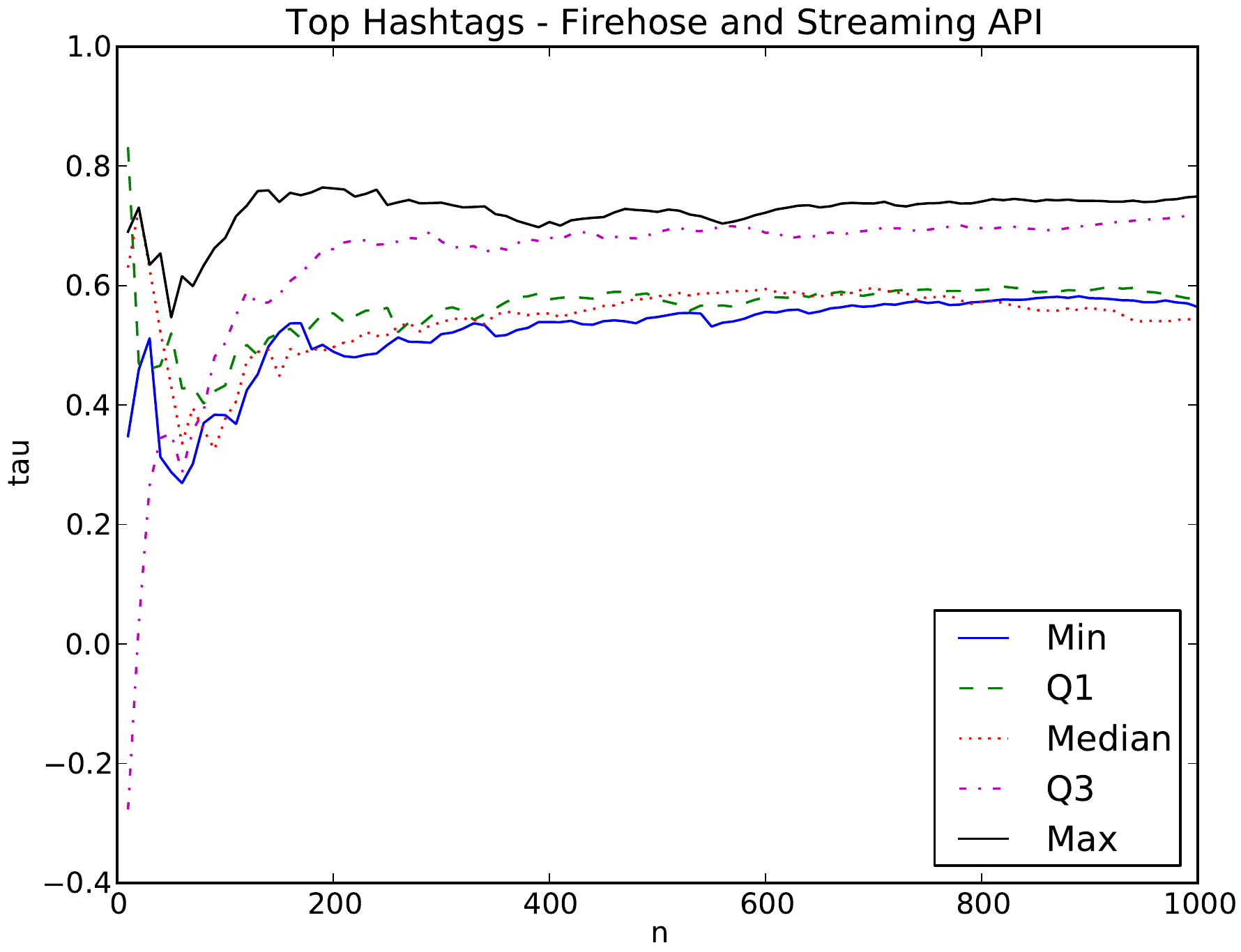}
  \caption{Relationship between $n$ - number of top hashtags, and the correlation coefficient, $\tau_\beta$.}
  \label{fig:taun}
  \vspace{-.1in}
\end{figure}

To understand the relationship between $n$ and the resulting correlation, $\tau_\beta$, we construct a chart showing the value of $\tau_\beta$ for $n$ between 10 and 1000 in steps of 10. To get an accurate representation of the differences in correlation at each level of Streaming coverage, we select five days with different levels of coverage as motivated by Figure~\ref{fig:covbox}: The minimum (December 27th), lower quartile (December 24th), median (December 29th), upper quartile (December 18th), and the maximum (December 19th). The results of this experiment are shown in Figure~\ref{fig:taun}. Here we see mixed results at small values of $n$, indicating that the Streaming data may not be good for finding the top hashtags. At larger values of $n$, we see that the Streaming API does a better job of estimating the top hashtags in the Firehose data.

\begin{figure}[t]
  \includegraphics[width=0.48\textwidth]{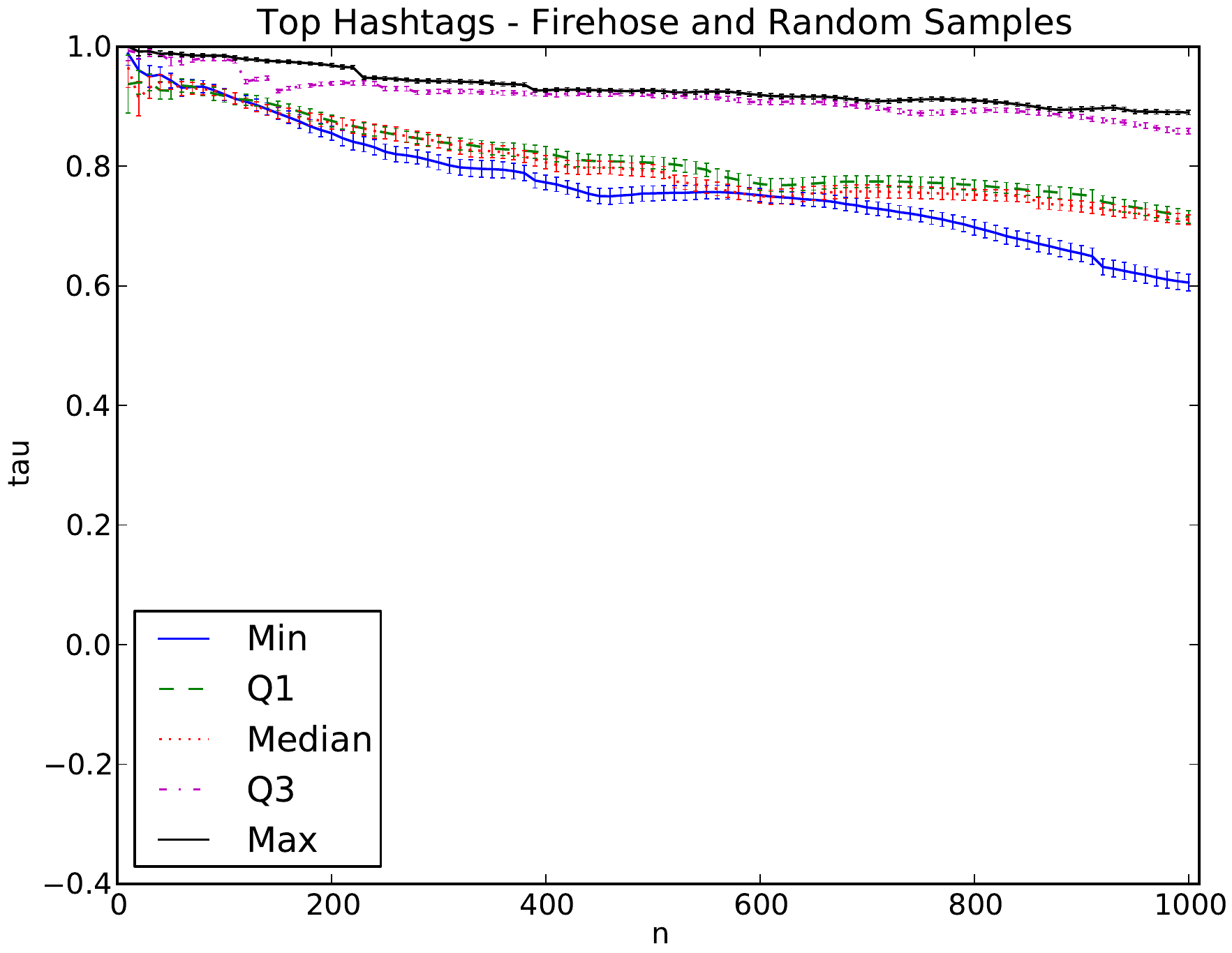}
  \caption{Random sampling of Firehose data. Relationship between $n$ - number of top hashtags, and $\tau_\beta$ - the correlation coefficient for different levels of coverage.}
  \label{fig:taunrandom}
\end{figure}

\subsubsection{Comparison with Random Samples}
After seeing the results from the previous section, we are left to wonder if the results are an artifact of using the Streaming API or if we could have obtained the same results by any random sampling. Would we obtain the same results with a random sample of equal size from the Firehose data, or does the Streaming API's filtering mechanism give us an advantage? To answer this question we repeat the experiments for each day in the previous section. This time, instead of using Streaming API data, we select tweets uniformly at random (without replacement) until we have amassed the same number of tweets as we collected from the Streaming API for that day. We repeat this process 100 times and obtain results as shown in Figure~\ref{fig:taunrandom}. Here we see that the levels of coverage in the random and Streaming data have comparable $\tau_\beta$ values for large $n$, however at smaller $n$ we see a much different picture. The random data gets very high $\tau_\beta$ scores for $n = 10$, showing a good capacity for finding the top hashtags in the dataset. The Streaming API data does not consistently find the top hashtags, in some cases revealing reverse correlation with the Firehose data at smaller $n$. This could be indicative of a filtering process in Twitter's Streaming API which causes a misrepresentation of top hashtags in the data.

\subsection{Topic Analysis}
Topic models are statistical models which discover topics in a corpus. Topic modeling is especially useful in large data, where it is too cumbersome to extract the topics manually. Due to the large volume of tweets published on Twitter, topic modeling has become central to many content-based studies using Twitter data \cite{kire09,Pozd-11,Hong12,Yin11,Chae-etal12}. We compare the topics drawn from the Streaming data with those drawn from the Firehose data using a widely-used topic modeling algorithm, latent Dirichlet allocation (LDA)~\cite{blei-etal03}. Latent Dirichlet allocation is an algorithm for the automated discovery of topics. LDA treats documents as a mixture of topics, and topics as a mixture of words. Each topic discovered by LDA is represented by a probability distribution which conveys the affinity for a given word to that particular topic. We analyze these distributions to understand the differences between the topics discovered in the two datasets. To get a sense of how the topics found in the Streaming data compare with those found with random samples, we compare with topics found by running LDA on random subsamples of the Firehose data.

\begin{figure*}[t]
     \begin{center}
        \subfigure[Min. $\mu = 0.024$,\newline $~~~~~~~~~~~~~~~~\sigma = 0.019.$]{
            \label{fig:first}
            \includegraphics[width=0.18\textwidth]{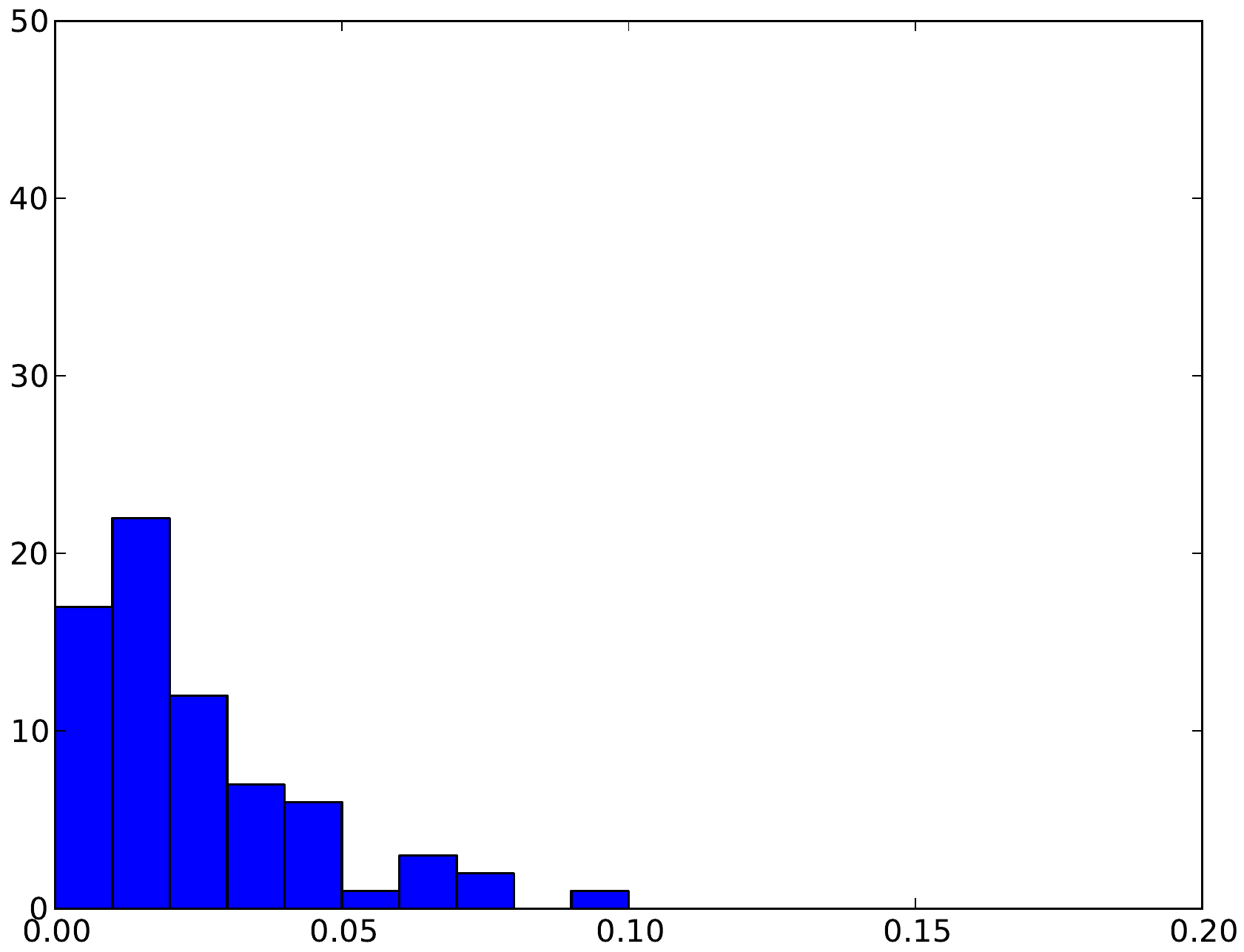}
        }%
        \subfigure[Q1. $\mu = 0.018$,\newline $~~~~~~~~~~~~~~~\sigma = 0.018.$]{%
           \label{fig:second}
           \includegraphics[width=0.18\textwidth]{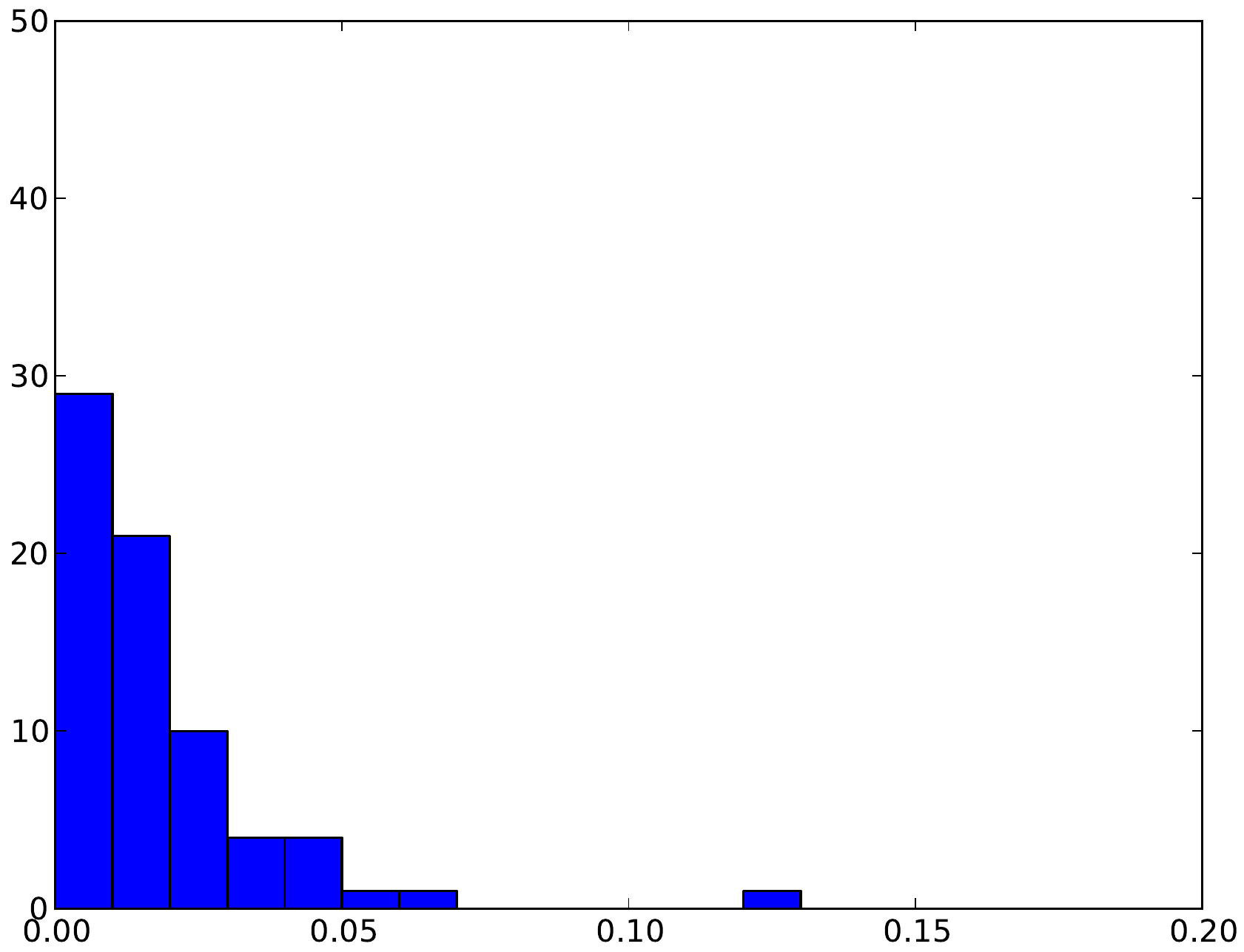}
        } %
        \subfigure[Median. $\mu = 0.018$,\newline $~~~~~~~~~~~~~~~~~~~~~\sigma = 0.020.$]{%
            \label{fig:third}
            \includegraphics[width=0.18\textwidth]{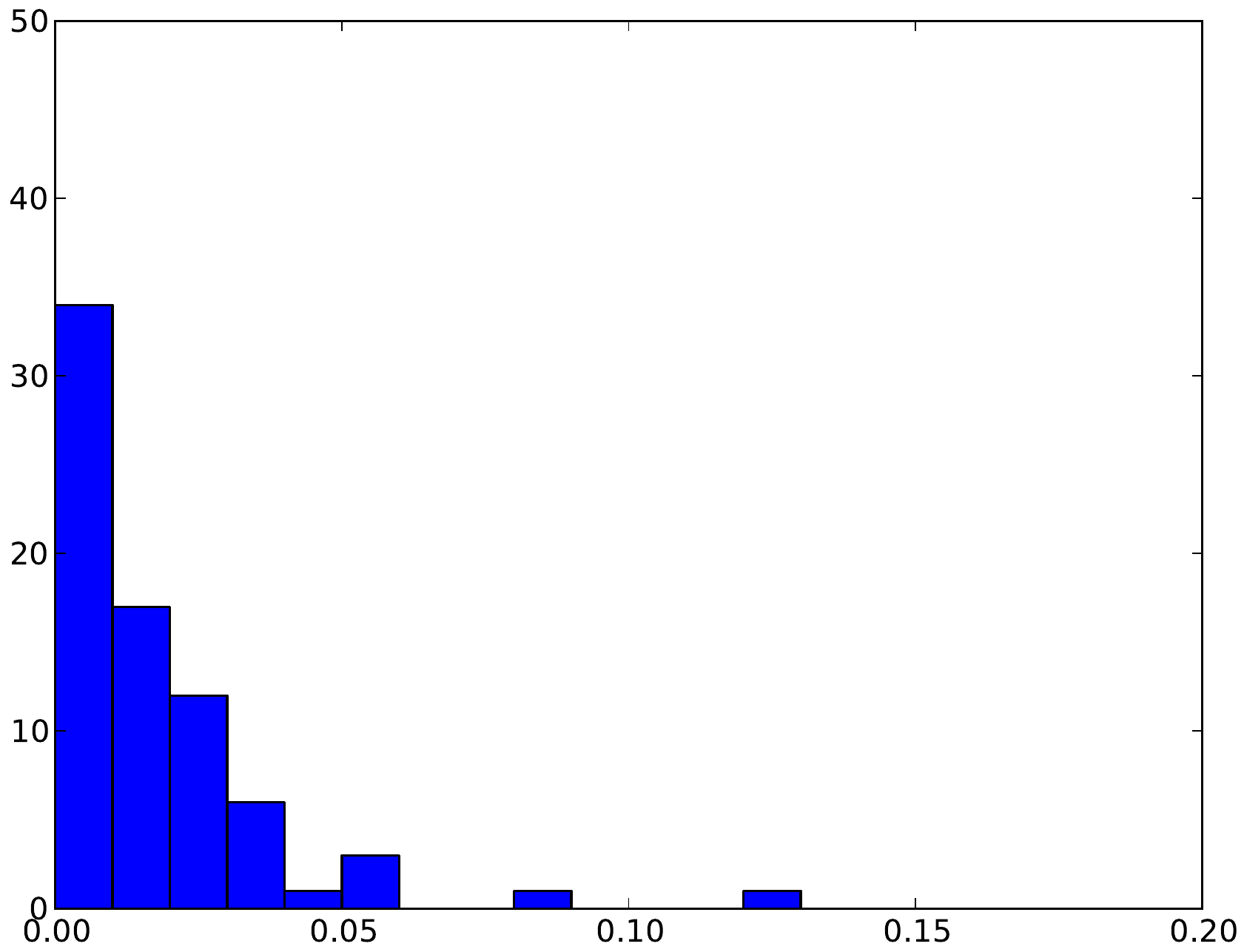}
        } %
        \subfigure[Q3. $\mu = 0.014$,\newline $~~~~~~~~~~~~~~\sigma = 0.016.$]{%
            \label{fig:fourth}
            \includegraphics[width=0.18\textwidth]{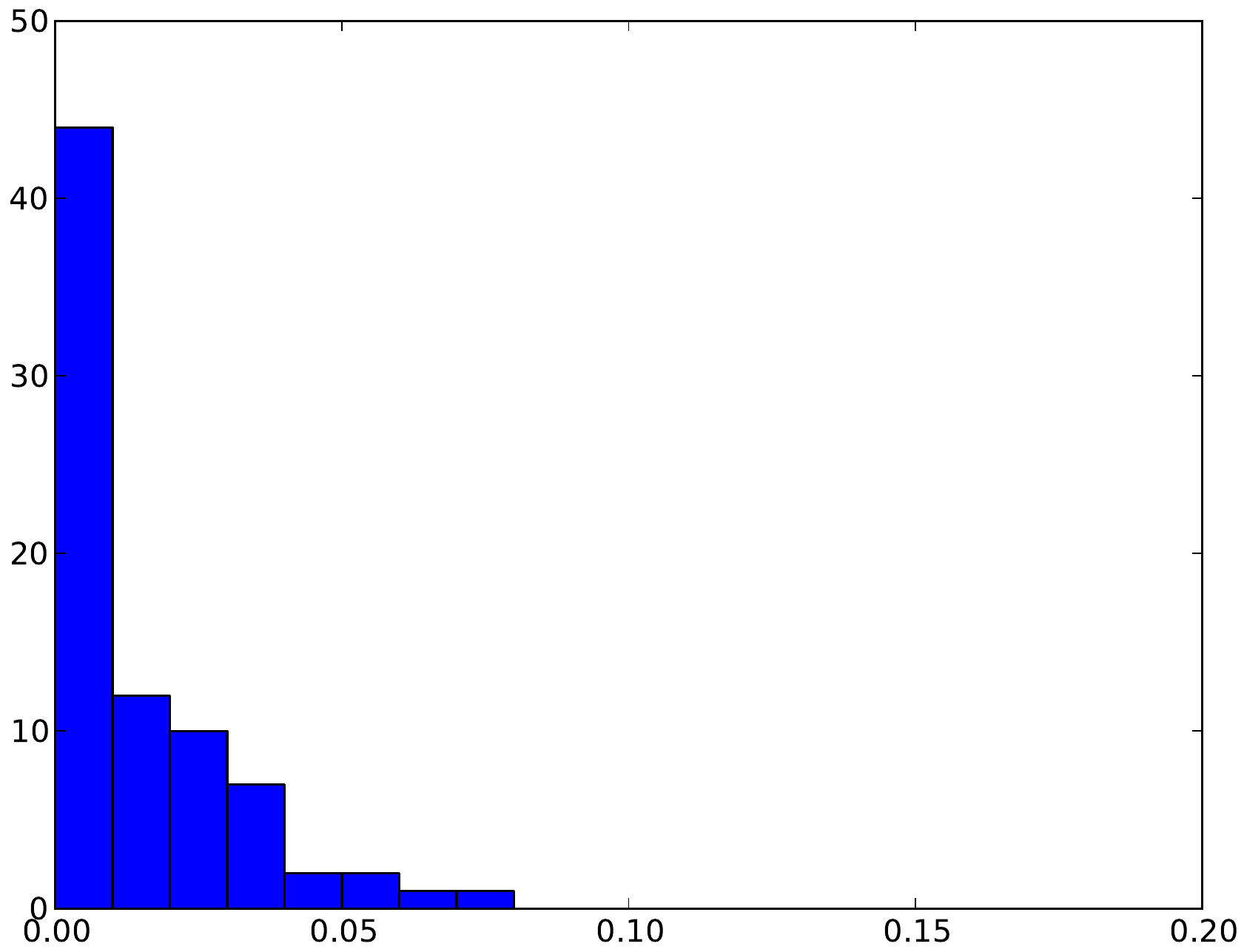}
        }%
        \subfigure[Max. $\mu = 0.016$,\newline$~~~~~~~~~~~~~~~~~\sigma = 0.018.$]{%
            \label{fig:fifth}
            \includegraphics[width=0.18\textwidth]{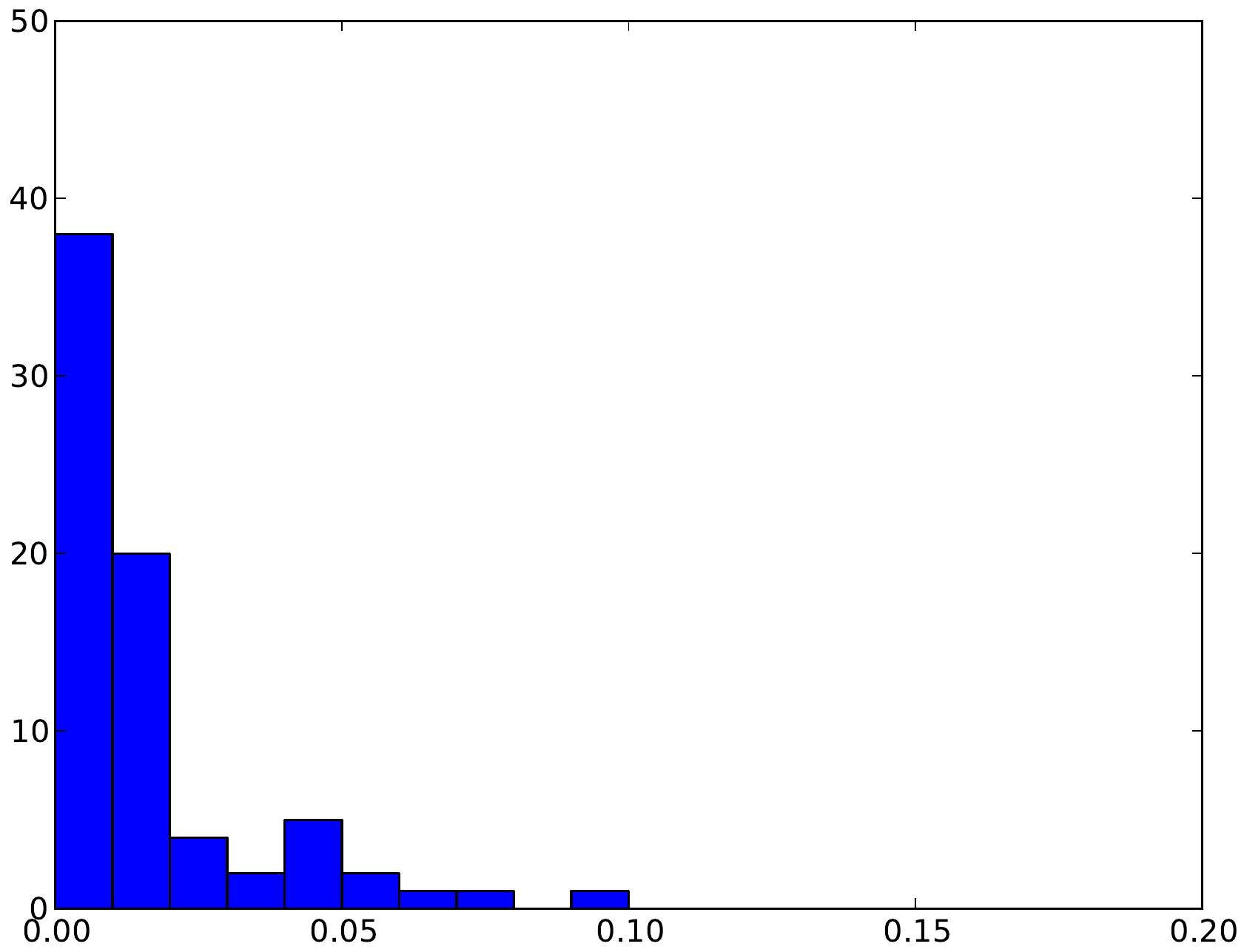}
        }%
    \end{center}
    \caption{%
        The Jensen-Shannon divergence of the matched topics at different levels of coverage. The x-axis is the binned divergence. No divergence was $>$ 0.15. The y-axis is the count of  each bin. $\mu$ is the average divergence of the matched topics, $\sigma$ is the standard deviation.
     }
   \label{fig:histograms}
\end{figure*}

\begin{figure*}[t]
     \begin{center}
        \subfigure[Min. $~S = 0.024$,\newline $~~~~~~~~~~~~~~~~~\hat{\mu} = 0.017$,\newline $~~~~~~~~~~~~~~~~~\hat{\sigma} = 0.002$,\newline $~~~~~~~~~~~~~~~~~z = 3.500$.]{
            \label{fig:firstdiff}
            \includegraphics[width=0.18\textwidth]{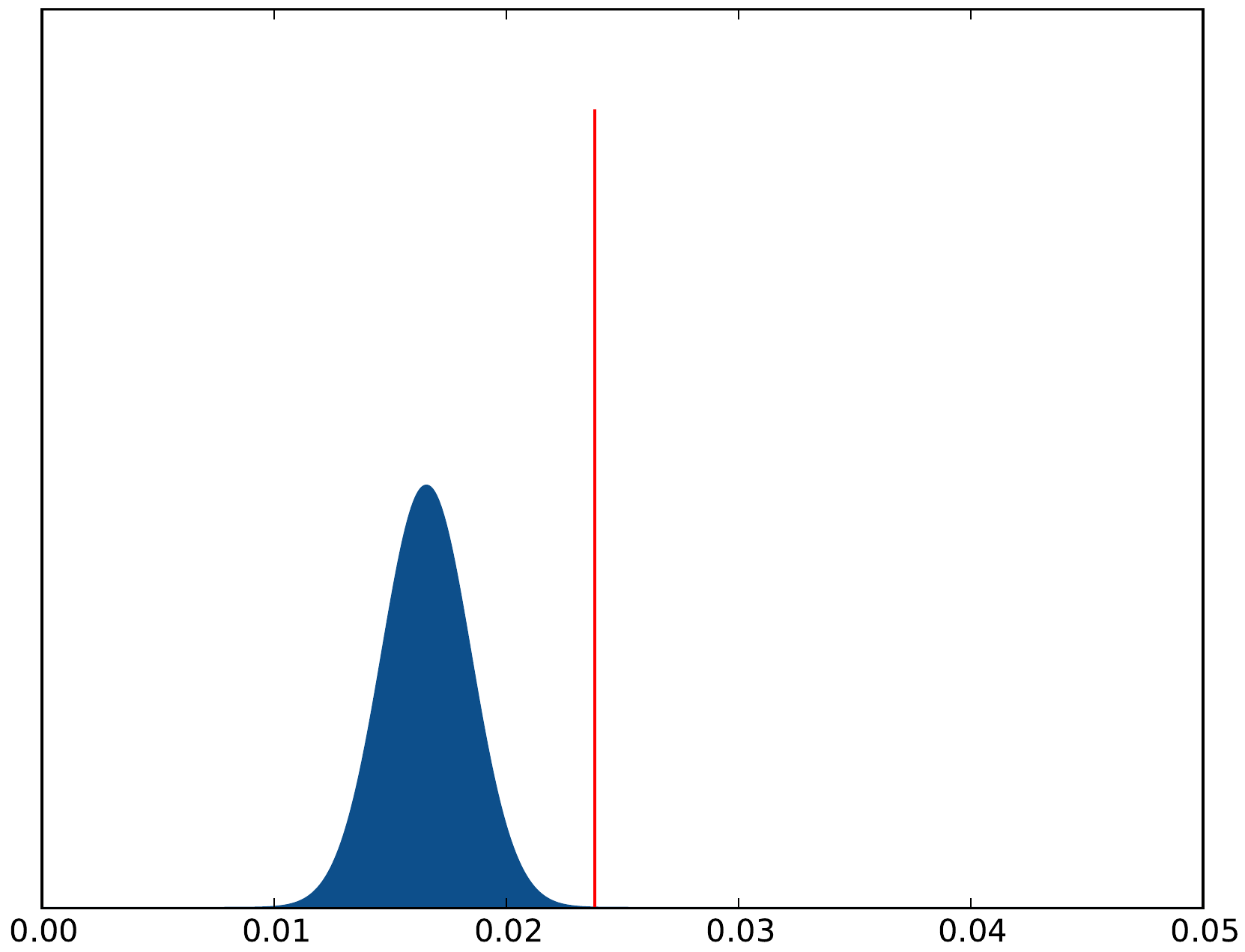}
        }%
        \subfigure[Q1. $~~S = 0.018$,\newline $~~~~~~~~~~~~~~~~~\hat{\mu} = 0.012$,\newline $~~~~~~~~~~~~~~~~~\hat{\sigma} = 0.001$,\newline $~~~~~~~~~~~~~~~~~z = 6.000$.]{%
           \label{fig:seconddiff}
           \includegraphics[width=0.18\textwidth]{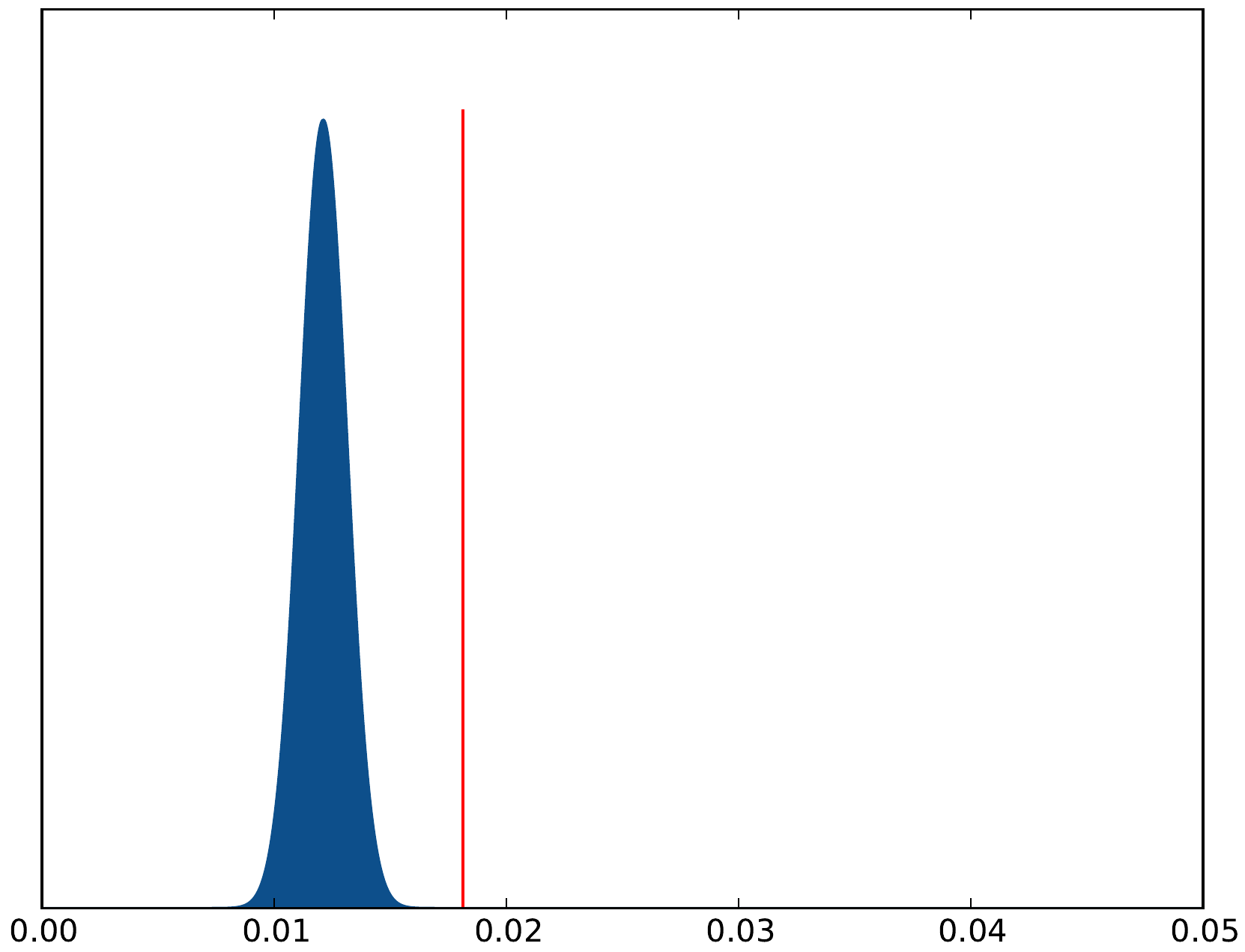}
        } %
        \subfigure[Median. $S = 0.018$,\newline $~~~~~~~~~~~~~~~~~~~~~\hat{\mu} = 0.013$,\newline $~~~~~~~~~~~~~~~~~~~~~\hat{\sigma} = 0.001$,\newline $~~~~~~~~~~~~~~~~~~~~~z = 5.000$.]{%
            \label{fig:thirddiff}
            \includegraphics[width=0.18\textwidth]{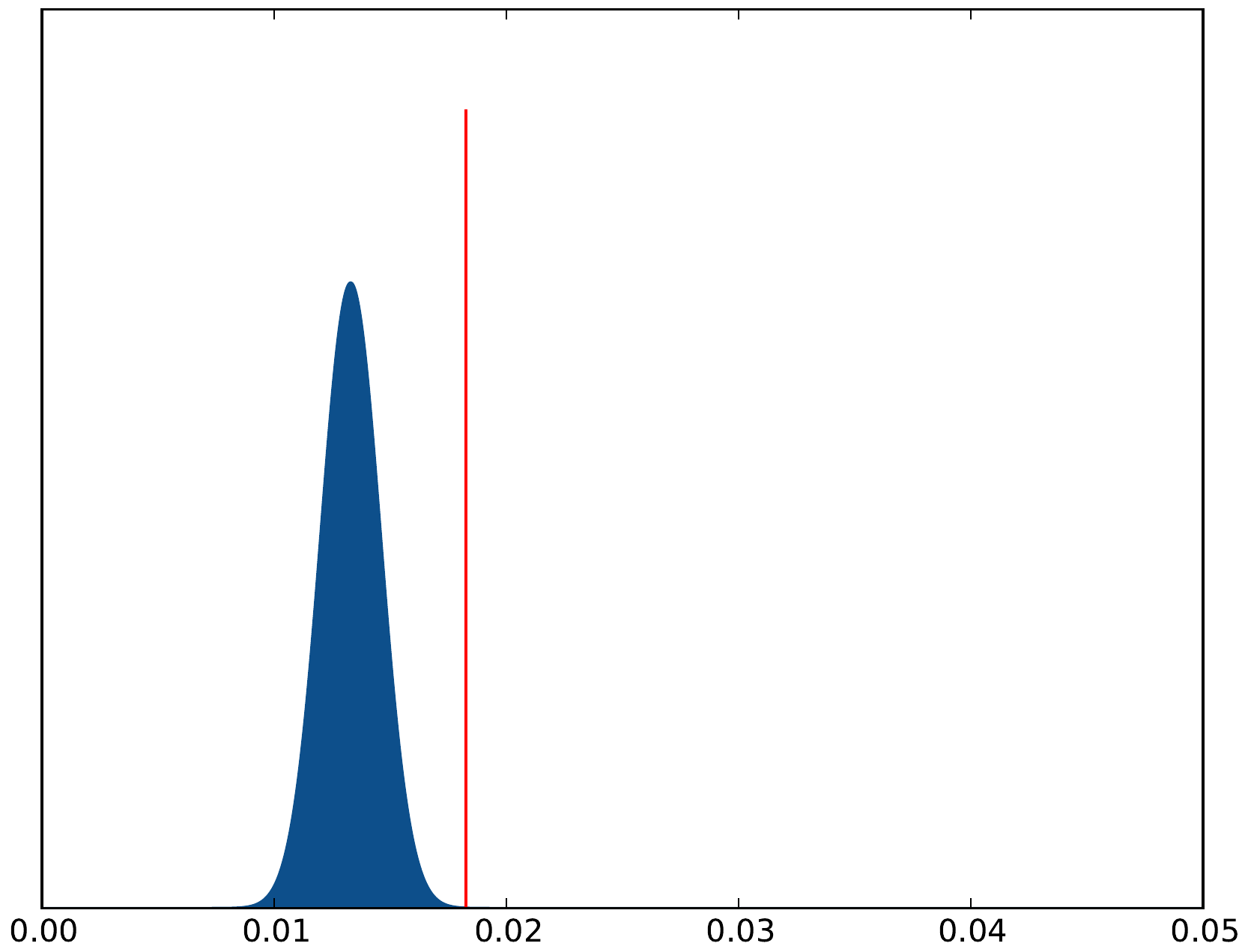}
        } %
        \subfigure[Q3. $~S = 0.014$,\newline $~~~~~~~~~~~~~~~~\hat{\mu} = 0.013$,\newline $~~~~~~~~~~~~~~~~\hat{\sigma} = 0.001$,\newline $~~~~~~~~~~~~~~~~z = 1.000$.]{%
            \label{fig:fourthdiff}
            \includegraphics[width=0.18\textwidth]{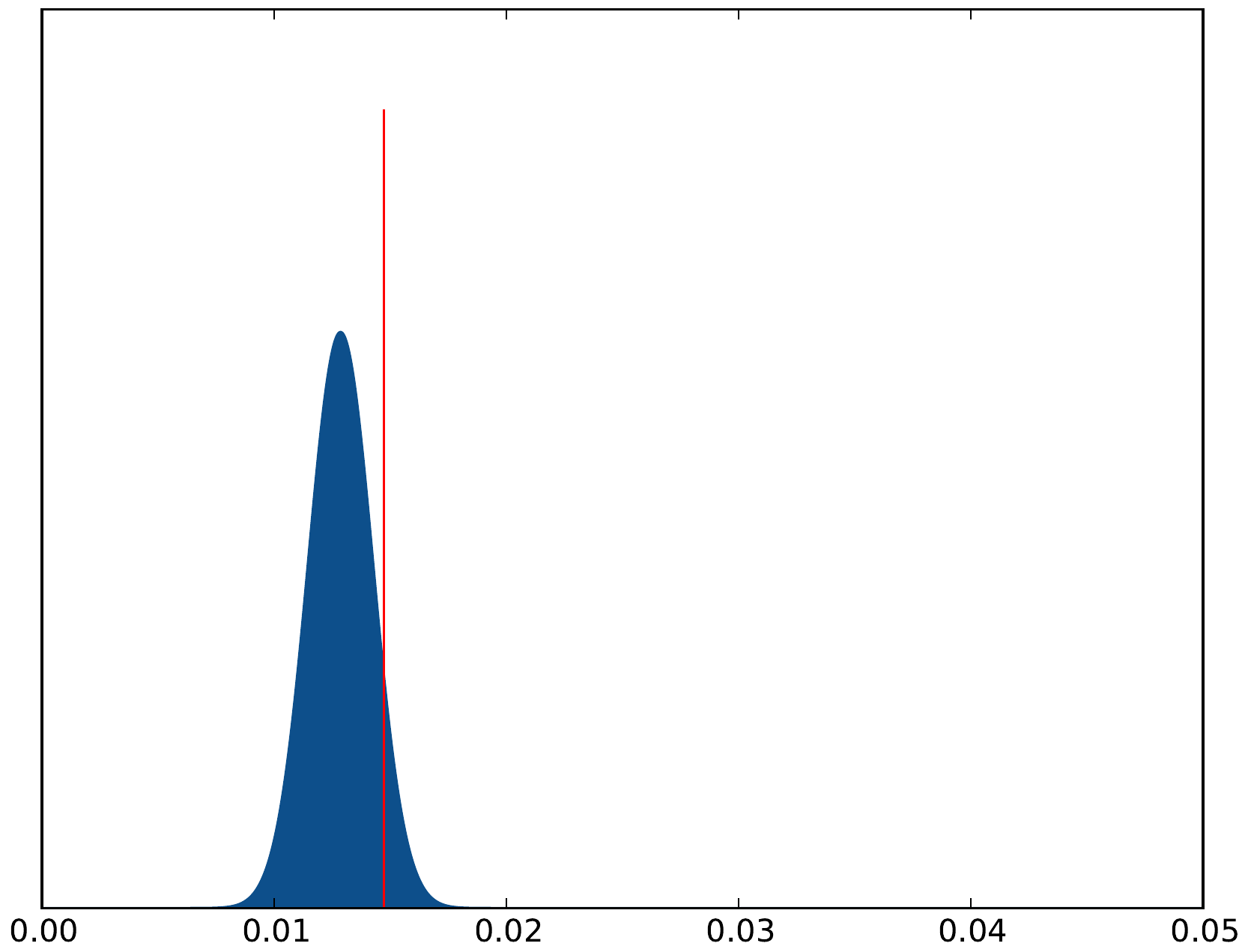}
        }%
        \subfigure[Max. $S = 0.016$,\newline $~~~~~~~~~~~~~~~~~\hat{\mu} = 0.013$,\newline $~~~~~~~~~~~~~~~~~\hat{\sigma} = 0.001$,\newline $~~~~~~~~~~~~~~~~~z = 3.000$.]{%
            \label{fig:fifthdiff}
            \includegraphics[width=0.18\textwidth]{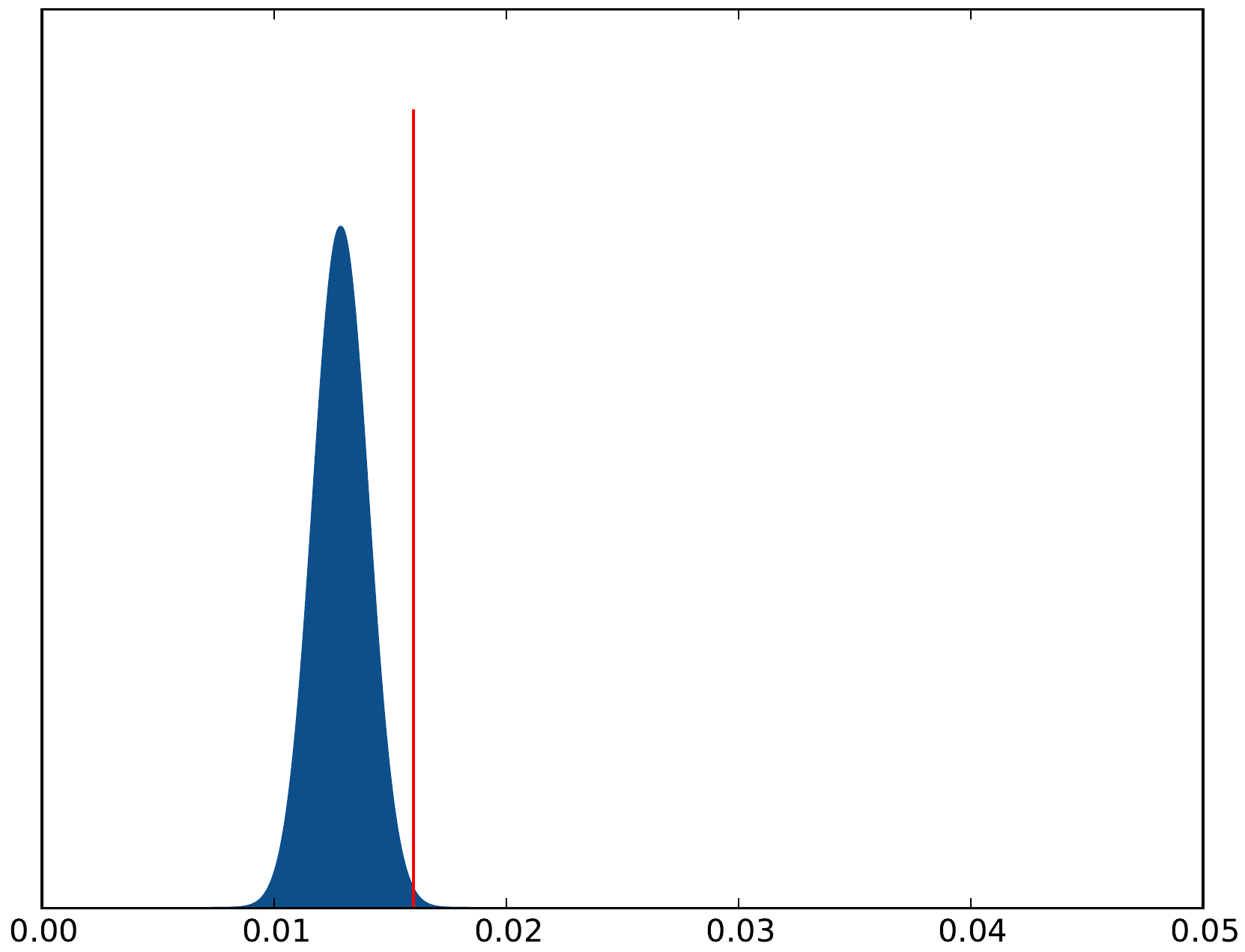}
        }%
    \end{center}
    \caption{%
        The distribution of average Jensen-Shannon divergences in the random data (blue curve), with the single average obtained through the Streaming data (red, vertical line). $z$ indicates the number of standard deviations the Streaming data is from the mean of the random samples.
     }
   \label{fig:pdfs}
\end{figure*}

\subsubsection{Topic Discovery}
Here we compare the topics generated using the Firehose corpus with those generated using the Streaming corpus. LDA takes, in addition to the corpus, three parameters as its input: $K$ - the number of topics, $\alpha$ - a hyperparameter for the Dirichlet prior topic distribution, and $\eta$ - a hyperparameter for the Dirichlet prior word distribution. Choosing optimal parameters is a very challenging problem, and is not the focus of this work. Instead we focus on the similarity of the results given by LDA using identical parameters on both the Streaming and Firehose corpus. We set $K = 100$ as suggested by~\cite{Duma88} and use priors of $\alpha = 50/K$, and $\eta = 0.01$. The software we used to discover the topics is the {\em gensim} software package~\cite{gensim}. To get an understanding of the topics discovered at each level of Streaming coverage, we select the same days as we did for the comparison of Kendall's $\tau$.

\subsubsection{Topic Comparison}
To understand the differences between the topics generated by LDA, we compute the distance in their probability distribution using the Jensen-Shannon divergence metric~\cite{Lin91}. Since LDA's topics have no implicit orderings we first must match them based upon the similarity of the words in the distribution. To do the matching we construct a weighted bipartite graph between the topics from the Streaming API and the Firehose. Treating each topic as a bag of words, we use the Jaccard score between the words in a Streaming topic $T^S_i$ and a Firehose topic $T^F_j$ as the weight of the edges in the graph,
\begin{equation}
  d(T^S_i, T^F_j) = \frac{|T^S_i \cap T^F_j|}{|T^S_i \cup T^F_j|}.
\end{equation}
After constructing the graph we use the maximum weight matching algorithm proposed in~\cite{Gali86} to find the best matches between topics from the Streaming and Firehose data. After making the ideal matches, we then compute the Jensen-Shannon divergence between the two topics. Treating each topic as a probability distribution, we compute this as follows:
\begin{equation}
JS(T^S_i || T^F_j) = \frac{1}{2}[KL(T^S_i || M) + KL(T^F_j || M)],
\end{equation}
where $M = \frac{1}{2}(T^S_i + T^F_j)$ and $KL$ is the Kullback-Liebler divergence~\cite{cove06}. We compute the Jensen-Shannon divergence for each matched pair and plot a histogram of the values in Figure~\ref{fig:histograms}. We see a trend of higher divergence with lower coverage, and lower divergence with higher coverage. This shows that decreased coverage in the Streaming data causes variance in the discovered topics.

\subsubsection{Comparison with Random Samples}
In order to get additional perspective on the accuracy of the topics discovered in the Streaming data, we compare the Streaming data with data sampled randomly from the Firehose, as we did earlier to compare the correlation. First, we compute the average of the Jensen-Shannon scores from the Streaming data in Figure~\ref{fig:histograms}, $S$. We then repeat this process for each of the 100 runs with random data, each run called $x_i$. Next, we use maximum-likelihood estimation~\cite{casella2001statistical} to estimate the parameters of the Gaussian distribution from which these points originate, $\hat{\mu} = \frac{1}{100} \sum_{i = 1}^{100} x_i$, and $\hat{\sigma} = \sqrt{\frac{1}{100}\sum_{i = 1}^{100}(x_i - \hat{\mu})^2}$. Finally, we compute the $z$-Score for $S$, $z = \frac{S - \hat{\mu}}{\hat{\sigma}}$. This score gives us a concrete measure of the difference between the Streaming API data and the random samples. Results of this experiment, including $z$-Scores are shown in Figure~\ref{fig:pdfs}. Nonetheless, we are still able to get topics from the Streaming API that are close to those found in random data with higher levels of coverage. A threshold of \emph{3-sigma} is often used in the literature to indicate extreme values~\cite[Section 6.3.1]{nist02}. With this threshold, we see that overall we are able to get significantly better topics with the random data than with the Streaming API on 4 of the 5 days.

\section{Network Measures}
Because Twitter is a social network, Twitter data can be analyzed with methods from Social Network Analysis~\cite{Wasserman1994} in addition to statistical measures. Possible 1-mode and 2-mode networks are: \emph{User $\times$ User} retweet networks, \emph{User $\times$ Hashtag} content networks, \emph{Hashtag $\times$ Hashtag} co-occurrence networks. For the purpose of this article we focus on \emph{User $\times$ User} retweet networks. Users who send tweets within a certain time period are the nodes in the network. Furthermore, users that are retweeted within this time period are also nodes in this network, regardless of the time their original tweet was tweeted.
The networks created by this procedure are directed and not symmetric by design, however, bi-directional links are possible in case $a \rightarrow b$ and $b \rightarrow a$. We ignore line weight created by multiple $a \rightarrow b$ retweets and self-loops (yes, some user retweet themselves). For the network metrics, the comparison is done on both the network, and the node levels. Networks are analyzed using ORA~\cite{ora2012}.

\subsection{Node-Level Measures}
The node-level comparison is accomplished by calculating measures at the user-level and comparing these results. %  (similar to previous sections). 
We calculate three different \emph{centrality measures} at the node level, two of which---Degree Centrality and Betweenness Centrality---were defined by Freeman as ``distinct intuitive conceptions of centrality''~\cite[p. 215]{Freeman1979}. Degree Centrality counts the number of neighbors in unweighted networks. In particular, we are interested in In-Degree Centrality as this reveals highly respected sources of information in the retweet network (where directed edges point to the source). Betweenness Centrality~\cite{Freeman1979} identifies brokerage positions in the Twitter networks that connect different communities with each other or funnel different information sources. Furthermore, we calculate the \emph{Potential Reach} which counts the number of nodes that are reachable in the network weighted with the path distance. In our Twitter networks this is equivalent to the inverse in-distance of reachable nodes~\cite{Sabidussi1966}. This approach results in a metric that finds sources of information (users) that potentially can reach many other nodes on short path distances. Before calculating these measures, we extract the main component and delete all other nodes (see next sub-section). In general, centrality measures are used to identify important nodes. Therefore, we calculate the number of top 10 and top 100 nodes that can be correctly identified with the Streaming data. Table~\ref{tab:centrality} shows the results for the average of 28 daily networks, the \emph{min-max} range, as well as the aggregated network including \emph{all} 28 days.
\begin{table}
  \caption{Average centrality measures for Twitter retweet networks for 28 daily networks. ``All'' is all 28 days together.}
  \renewcommand{\arraystretch}{1.2}
  \begin{tabular}{| >{\centering\arraybackslash}p{2.5cm} | >{\centering\arraybackslash}p{1.1cm} | >{\centering\arraybackslash}p{1.9cm} | >{\centering\arraybackslash}p{1.1cm} |}
  \hline
  Measure & $k = $ & Top$-k$ \emph{(min-max)} & All \\
  \hline
  In-Degree & 10 & 4.21 (0--9) & 4 \\
  \hline
  In-Degree & 100 & 53.4 (36--82) & 73 \\
  \hline
  Potential Reach & 100 & 59.2 (32--83) & 80\\
  \hline
  Betweenness & 100 & 54.8 (41--81) & 55 \\
  \hline
  \hline
  \end{tabular}
  \label{tab:centrality}
\end{table}

Although, we know from previous studies ~\cite{Borg06} that there is a very low likelihood that the ranking will be correct when handling networks with missing data, the accuracy of the daily results is not very satisfying. When we look at the results of the individual days, we can see that the matches have, once again, a broad range as a function of the data coverage rate. In~\cite{Borg06} the authors argue that network measures are stable for denser networks. Twitter data, being very sparse, causes the network metrics' accuracy to be rather low in the case when the data sub-sample is smaller. However, identifying $\sim$50\% key-players correctly for a single day is reasonable, and accuracy can be increased by using longer observation periods. Even more, the Potential Reach metrics are quite stable for some days in the aggregated data.

\subsection{Network-Level Measures}
We complement our node-level analysis by comparing various metrics at the network level. These metrics are reported in Table~\ref{tab:network1} and are calculated as follows. Since retweet networks create a lot of small disconnected components, we focus only on the size of the largest component. The size of the main component and the fact that all smaller components contain less than 1\% of the nodes justify our focus on the main component for this data. Therefore, we reduce the networks to their largest component before we proceed with the calculations. To describe the structure of the retweet networks we calculate the clustering coefficient, a measure for local density~\cite{Watts1998}. We do not take all possible triads of directed networks into account, but treat the networks as undirected when calculating the clustering coefficient. $D_{in}>0$ shows the proportion of nodes in the largest component that are retweeted and $max(D_{in})$ shows the value of the highest unscaled In-Degree value, i.e., number of unique users retweeting the same single user. The final three lines of Table~\ref{tab:network1} are network centralization indexes based on the node-level measures that have been introduced in the previous paragraph. Freeman~\cite{Freeman1979} describes the centralization $C_X$ of a network for any given metric as the difference of the value $C_X(p*)$ of the most central node to all other node values compared to the maximum possible difference:
\begin{equation}
C_X =  \frac{\sum{i=1}{n}{[C_X(p*)-C_X(p_i)]}}{max \sum{i=1}{n}{[C_X(p*)-C_X(p_i)]}}
\end{equation}
High centralization indicates a network with some nodes having very high node-level values and many nodes with low values while low centralization is the result of evenly distributed node-level measures.

We do not discuss all details of the individual results but focus on the differences between the two data sources. First, the coverage of nodes and links is similar to the coverage of tweets. This is a good indicator that the sub-sample is not biased to the specific Twitter user (e.g. high activity). The smaller proportion of nodes with non-zero In-Degree for the Firehose shows us that the larger number of nodes includes many more peripheral nodes. A low Clustering Coefficient implies that networks are hierarchical rather than interacting communities. Even though the centralization indexes are rather similar, there is one very interesting result when looking at the individual days: The range of values is much higher for the Streaming data as a result of the high coverage fluctuation. Further research will analyze whether we can use network metrics to better estimate how sufficient the sampled Streaming data is.

\begin{table}
  \centering
  \caption{Comparison of Network-Level Social Network Analysis Metrics.}
  \renewcommand{\arraystretch}{1.2}
  \small \begin{tabular}{| >{\centering\arraybackslash}p{2.1cm} | >{\centering\arraybackslash}p{1cm} | >{\centering\arraybackslash}p{1cm} | >{\centering\arraybackslash}p{1cm} | >{\centering\arraybackslash}p{1cm} |}
  \hline
   & \multicolumn{2}{c|}{Firehose} & \multicolumn{2}{c|}{Streaming API} \\
  \hline
  Metrics & avg.day & 28 days & avg.day & 28 days \\
  \hline
nodes & 6,590 & 73,719  & 2,466 \small(37.4\%) &  30,894 \small(41.9\%)\\
links & 10,173 & 204,022 & 3,667 \small(36.0\%) & 76,750 \small(37.6\%) \\
$D_{in}>0$ & 25.1\% & 19.3\% & 32.4\% & 20.5\% \\
$max(D_{in})$ & 341 & 2,956 & 167.3 & 1,252 \\
 \hline
main comp. & 5,609 & 70,383 &  2,069 & 28,701 \\
main comp. \% & 84.6\% & 95.5\% & 82.5\% & 92.9\% \\
 \hline
Clust.Coef. & 0.029 & 0.053 & 0.033 & 0.050 \\
$DC_{in}$ Centr. & 0.059 & 0.042 & 0.085 & 0.043 \\
$BC$ Centr. &  0.010 & 0.053 & 0.010 & 0.050 \\
$PReach$ Centr. & 0.130 & 0.240 & 0.156 & 0.205 \\
  \hline
  \hline
  \end{tabular}
  \label{tab:network1}
\end{table}

\section{Geographic Measures}
The final facet of the Twitter data we compare is the geolocation of the tweets. Geolocation is an important part of a tweet, and the study of the location of content and users is currently an active area of research~\cite{Chen10,Waka11}. We study how the geographic distribution of the geolocated tweets is affected by the sampling performed by the Streaming API.

The number of geotagged tweets is low, with only 16,739 geotagged tweets in the Streaming data (3.17\%) and 18,579 in the Firehose data (1.45\%). We notice that despite the difference in tweets collected on the whole we get 90.10\% coverage of geotagged tweets. We start by grouping the locations of tweets by continent and can find a strong Asian bias due to the boundary box we used to collect the data from both sources, shown in Table~\ref{tab:params}. To better understand the distribution of geotagged tweets we repeat the same process, this time excluding tweets originating in the boundary box set in the parameters. After removing these tweets, more than 90\% of geotagged Tweets from both sources are excluded from the data and the Streaming coverage level is reduced to 39.19\%. The distribution of tweets by continent is shown in Table~\ref{tab:geocontnobox}. Here we see a more even representation of the tweets' locations in Asia and North America. 

\begin{table}
  \caption{Geotagged Tweet Location by Continent. Excluding boundary box from parameters.}
  \renewcommand{\arraystretch}{1.2}
  \small \begin{tabular}{| c | r | r | r |}
  \hline
  Continent & Firehose & Streaming & Error \\
  \hline
  Africa & 156 \small(5.74\%) & 33 \small(3.10\%) & -2.64\% \\
  Antarctica & 0 \small(0.00\%) & 0 \small(0.00\%) & $\pm$0.00\% \\
  Asia & 932 \small(34.26\%) & 321 \small(30.11\%) & -4.15\% \\
  Europe & 300 \small(11.03\%) & 139 \small(13.04\%) & +2.01\% \\
  Mid-Ocean & 765 \small(28.12\%) & 295 \small(27.67\%) & -0.45\% \\
  N. America & 607 \small(22.32\%) & 293 \small(27.49\%) & +5.17\% \\
  Oceania & 54 \small(1.98\%) & 15 \small(1.41\%) & -0.57\% \\
  S. America & 3 \small(0.11\%) & 2 \small(0.19\%) & +0.08\% \\
  \hline
  Total & 2720 \small(100.00\%) & 1066 \small(100.00\%) & $\pm$0.00\% \\
  \hline
  \hline
  \end{tabular}
  \vspace{-.2in}
  \label{tab:geocontnobox}
\end{table}

\section{Conclusion and Future Work}
In this work we ask whether data obtained through Twitter's sampled Streaming API is a sufficient representation of activity on Twitter as a whole. To answer this question we collected data with exactly the same parameters from both the free, but limited, Streaming API and the unlimited, but costly, Firehose. We provide a methodology for comparing the two multifaceted sets of data and results of our analysis.

We started our analysis by understanding the coverage of the Streaming API data, finding that when the number of tweets matching the set of parameters increases, the Streaming API's coverage is reduced. One way to mitigate this might be to create more specific parameter sets with different users, bounding boxes, and keywords. This way we might be able to extract more data from the Streaming API.

Next, we studied the statistical differences between the two datasets. We used a common correlation coefficient to understand the differences between the top $n$ hashtags in the two datasets. We find that the Streaming API data estimates the top hashtags for a large $n$ well, but is often misleading when $n$ is small. 
We also employed LDA to extract topics from the text. We compare the probability distribution of the words from the most closely-matched topics and find that they are most similar when the coverage of the Streaming API is greatest. That is, topical analysis is most accurate when we get more data from the Streaming API. 

The Streaming API provides just one example of how sampling Twitter data affects measures. We leverage the Firehose data to get additional samples to better understand the results from the Streaming API. In both of the above experiments we compare the Streaming data with 100 datasets sampled randomly from the Firehose data. We compare the statistical properties to find that the Streaming API performs worse than randomly sampled data, especially at low coverage. We find that in the case of top hashtag analysis, the Streaming API sometimes reveals negative correlation in the top hashtags, while the randomly sampled data exhibits very high positive correlation with the Firehose data. In the case of LDA we find a significant increase in the accuracy of LDA with the randomly sampled data over the data from the Streaming API. Both of these results indicate some bias in the way that the Streaming API provides data to the user.

By analyzing retweet \emph{User $\times$ User} networks we were able to show that we can identify, on average, 50--60\% of the top 100 key-players when creating the networks based on one day of Streaming API data. Aggregating some days of data can increase the accuracy substantially. For network level measures, first in-depth analysis revealed interesting correlation between network centralization indexes and the proportion of data covered by the Streaming API.

Finally, we inspect the properties of the geotagged tweets from both sources. Surprisingly, we find that the Streaming API almost returns the complete set of the geotagged tweets despite sampling. We attribute this to the geographic boundary box. Although the number of geotagged tweets is still very small in general ($\sim$1\%), researchers using this information can be confident that they work with an almost complete sample of Twitter data when geographic boundary boxes are used for data collection. When we remove the tweets collected this way, we see a much larger disparity in the tweets from both datasets. Even with this disparity, we see a similar distribution based on continent.

Overall, we find that the results of using the Streaming API depend strongly on the coverage and the type of analysis that the researcher wishes to perform. This leads to the next question concerning the estimation of how much data we actually get in a certain time period. We suggest that we found first evidence in different types of analysis that can help us to estimate the Streaming API coverage. Uncovering the nuances of the Streaming API will help researchers, business analysts, and governmental institutions to better ground their scientific results based on Twitter data.

Looking forward, we hope to find methods to compensate for the biases in the Streaming API to provide a more accurate picture of Twitter activity to researchers. Provided further access to Twitter's Firehose, we will determine whether the methodology presented here will yield similar results for Twitter data collected from other domains, such as natural disaster, protest, and elections.

\section{Acknowledgements}
This work is sponsored in part by Office of Naval Research grants N000141010091 and N000141110527. We thank Lei Tang for thoughtful advice and discussion throughout the development of this work.
\bibliographystyle{aaai}
\bibliography{references}

\end{document}